\documentclass[12pt,showpacs,preprintnumbers,aps]{revtex4}
\usepackage {amsfonts}
\usepackage {graphicx}
\usepackage {epsfig}
\usepackage{amssymb}
\usepackage{amsmath}
\usepackage[english]{babel}

\begin{document}

\title{Search for the neutron EDM and time reversal symmetry violation in noncentrosymmetric crystals}

\author{V.G. Baryshevsky}
\address{Research Institute for Nuclear Problems, 11 Bobryiskaya str.,
\\ 220050, Minsk, Belarus, e-mail:bar@inp.minsk.by}

\begin{abstract}
CP-violating spin rotation and spin dichroism in
noncentrosymmetric crystals is discussed.
\end{abstract}


\maketitle

The origin of CP-symmetry violation (where C is a charge
conjugation and P is a spatial inversion) is of a great interest
since its discovery in the decay o neutral K-mesons about 40 years
ago.
CP-violation leads in turn to the violation of the time reversal
symmetry (T) through the CPT invariance (CPT-theorem).
Existence of nonzero neutron EDM requires violation of both P and
T invariance.
Different theories of CP violation give widely varying predictions
for a neutron EDM.
This is the reason now for discussion of two types of experiments:

\noindent 1. study of spin precession frequency for ultracold
neutrons in magnetic and electric fields \cite{1,2,3}

\noindent 2. study of neutron diffraction in a noncentrosymmetric
crystal \cite{VG,Bar,Forte1983,Forte1989,LIYAF,LIYAF1}.

Experiments studying neutron diffraction in the noncentrosymmetric
crystal are of particular interest because they provide to get
limits for the T-violating term in the amplitude of coherent
elastic scattering at a non-zero angle
$f_T(\vec{k},\vec{k}^{\prime}) \sim
\vec{\sigma}(\vec{k}^{\prime}-\vec{k})$, where $\vec{k}$ is the
wave vector of the neutron incident on the crystal,
$\vec{k}^{\prime}$ is the wave vector of the scattered neutron,
$\vec{\sigma}=(\sigma_x,\sigma_y,\sigma)$ are the Pauli matrices,
which describe neutron spin.
According to \cite{VG,Bar} the amplitude
$f_T(\vec{k},\vec{k}^{\prime})$ contributes to the index of
refraction for neutrons in the noncentrosymmetric  crystal and
arouses T-violating spin rotation and spin dichroism even for
neutrons, for which condition of Bragg diffraction is not
fulfilled.
Requirements for crystal quality in this case appear less strict
\cite{VG,Bar} in comparison with those necessary for observation
of spin rotation and spin dichroism in conditions of dynamical
diffraction.
As a result, even observation of spin rotation and spin dichroism
for neutrons in the range of P-resonances appears possible
\cite{VG,Bar}.

Experimental methods for study of neutron spin rotation in
crystals have been developed recent years.
Application of these methods can significantly increase
sensitivity of EDM measurement for neutrons moving in
noncentrosymmetric crystals \cite{LIYAF}.
The limit for the neutron EDM in such experiments is expected
about $d_n \leq 10^{-27} \textrm{e} \cdot \textrm{cm}$.

According to \cite{VG,Bar} in such experiments the T-odd
scattering amplitude $f_T (\vec{k}^{\prime},\vec{k})$, which is
caused by T-odd nuclear interactions, also can be measured.
Therefore, increasing of the experiment sensitivity makes possible
more accurate measurement of $f_T (\vec{k}^{\prime},\vec{k})$,
which, in its turn, provides more strict limits for the
CP-violating constant of nucleon-nucleon interactions.

\section {Neutron spin rotation and spin dichroism in media}

Neutron spin rotation and spin dichroism in media caused by parity
(P) violation and possible time (T) noninvariance under neutron
interaction with nuclei are described by the refractive index of
neutrons in media
\begin{equation}
\hat N=1+\frac{2\pi \rho }{k^2}\hat f(0), \label{2.1}
\end{equation}
where $k$ is the neutron wave number, $\rho $ is the scatters
density, $\hat f(0)$ is the coherent elastic forward scattering
amplitude of neutrons on a nucleus.

To find the refraction index for neutrons in the
noncentrosymmetric crystal let us consider the Schrodinger
equation describing propagation of a coherent neutron wave $\Psi$
in a crystal \cite{VG,Bar}:
\begin{equation}
\left( k^2-k_0^2\right) \Psi \left( \vec k\right) +\sum_{\vec \tau }\frac{2m%
}{\hbar ^2}\hat U_{eff}\left( \vec \tau \right) \Psi \left( \vec
k-2\pi \vec \tau \right) =0, \label{2.2}
\end{equation}
where $k_{0}$ is the neutron wave number in vacuum, $k$ is the
neutron wave number in a crystal. The Fourier transform of neutron
effective potential energy in a crystal is

\begin{equation}
\hat U_{eff}\left( \vec \tau \right) =-\frac{2\pi \hbar
^2}{mV_0}\hat F\left( \vec \tau \right) ; \hat F\left( \vec \tau
\right) =\sum_j\hat f_j\left( \vec \tau \right) e^{-w_j\left( \vec
\tau \right) }e^{-i2\pi \vec \tau \vec R_j}, \label{2.3}
\end{equation}
$\hat F\left( \vec \tau \right) $ is the amplitude of neutron
coherent scattering on crystal unit cell in the direction $\vec
k^{^{\prime }}=\vec k+2\pi \vec \tau $ , where $2\pi \vec \tau $
is the vector of crystal reciprocal lattice,
$\hat f_j\left( \vec \tau \right) = Re \hat f_j\left( \vec \tau
\right) + i Im \hat f_j\left( \vec \tau \right)$ is the amplitude
$\hat f_j\left( \vec k^{^{\prime }},\vec k\right)$ of coherent
elastic scattering  from a j-type nucleus in the direction of
$\vec k^{^{\prime }}=\vec k+2\pi \vec \tau $,
$Re \hat f_j\left( \vec \tau \right)$ is the real part of the
amplitude,
$Im \hat f_j\left( \vec \tau \right)$ is its imaginary part,
$\vec R_j$ is the coordinate of the $j$ nucleus in the unit cell,
$e^{-w_j\left( \vec \tau \right) }$ is the Debay-Waller factor.

%---------- page 510
For a non-polarized nucleus
\begin{equation}
\hat f_j\left( \vec k^{^{\prime }},\vec k\right) = A_j +
f_{so}^j \vec{\sigma} [\vec{k} \times \vec{k}^{\prime}] +
C_j \vec{\sigma} \vec{\nu}_1 + C_j^{\prime} \vec{\sigma}
\vec{\nu}_2
 \label{JP5}
\end{equation}
where $A_j$ is the spin-independent part of the amplitude of
neutron scattering by the nucleus due to strong interaction,
$f_{so}^j \vec{\sigma} [\vec{k} \times \vec{k}^{\prime}]$ is the
spin-orbit scattering amplitude of a neutron by a nucleus,
$f_{so}^j = f_{son}^j + f_{Sch}^j$, where the amplitudes
$f_{son}^j$ and $f_{Sch}^j$ describe nuclear and Schwinger
interactions, respectively, $C_j \vec{\sigma} \vec{\nu}_1$ is the
P-violating T-invariant scattering amplitude, $C_j^{\prime}
\vec{\sigma} \vec{\nu}_2$ is the T- and P-violating scattering
amplitude, and the vectors $\vec{\nu}_1$ and $\vec{\nu}_2$ are
\begin{equation}
\vec{\nu}_1 = \frac{\vec{k}^{\prime} + \vec{k}}{|\vec{k}^{\prime}
+ \vec{k}|}
\vec{\nu}_2 = \frac{\vec{k}^{\prime} - \vec{k}}{|\vec{k}^{\prime}
- \vec{k}|}
\end{equation}

%-----

The system of homogeneous equations (\ref{2.2}) permits to
determine the dependence of $k$ on $k_0$, i.e. to determine the
neutron refractive index in a crystal.

It is well-known that the area $\Delta \vartheta $ of neutron
incidence angle on a crystal in which strong diffraction is
observed (when the diffracted wave amplitude is comparable with
the amplitude of incident wave) is very small $\Delta \vartheta
\sim 10^{-5}\div 10^{-6}$ rad even for thermal neutrons.

Outside this narrow angle area the diffracted wave amplitude is
small and the system of equations (\ref{2.2}) can be analyzed
according to the perturbation theory. As a result, we have

\begin{equation}
\hat N=\frac k{k_0}\simeq 1+\frac 12\hat g\left( 0\right)
+\frac{\hat g\left( -\vec \tau \right) \hat g\left( \vec \tau
\right) }{2\alpha _B}, \label{2.4}
\end{equation}

where

\[
\hat g\left( \vec \tau \right) =-\frac{2m}{\hbar ^2k_0^2}\hat
U_{eff}\left( \vec \tau \right) ,\alpha _B=\frac{2\pi \vec \tau
\left( 2\pi \vec \tau +2\vec k_0\right) }{k_0^2}
\]

As we see, the correction to the neutron refractive index in
crystals contains the scattering amplitude at the non-zero angle
$\hat f\left( \vec \tau \right) $.

According to \cite{VG,Bar} the expression for $\hat g\left( \vec
\tau \right) $ can be written as a sum

\begin{equation}
\hat g\left( \vec \tau \right) =\hat g_s\left( \vec \tau \right)
+\hat g_{so}\left( \vec \tau \right) +\hat g_w\left( \vec \tau
\right) , \label{2.5}
\end{equation}
where $\hat g_s\left( \vec \tau \right) $ is proportional to the
amplitude of scattering off the nucleus $\hat f_s\left( \vec \tau
\right) $ due to strong interactions without consideration
spin-orbital interactions; $\hat g_w\left(
\vec \tau \right) $ is proportional to the P and T-violating amplitude; $%
\hat g_{so}=\hat g_{son}+\hat g_{shw}$ , $\hat g_{son}$ is
proportional to the amplitude of spin-orbital neutron scattering
on the nucleus due to nuclear forces $\hat f_{son}\left( \vec \tau
\right) =f_{son}\vec \sigma \left[ \vec k\times 2\pi \vec \tau
\right] $ , $\hat g_{shw}$ is proportional to the amplitude of
spin orbital neutron scattering on the nucleus due to neutron
magnetic momentum interaction with the nucleus electric field (the
so-called Schwinger scattering):

\begin{equation}
\hat g_{shw}=i\frac{2m}{V_0\hbar ^2k_0^2}\frac{\mu \hbar
}{mc}\sum_j\Phi _j\left( \vec \tau \right) e^{-w_j\left( \vec \tau
\right) }\vec \sigma \left[ \vec k\times 2\pi \vec \tau \right]
e^{-i2\pi \vec \tau \vec \tau _j}, \label{2.6}
\end{equation}
$\mu $ is the magnetic neutron momentum, $\Phi _j\left( \vec \tau
\right)$ is the Fourier component of electrostatic potential
induced by nucleus $j$.

From equations (\ref{JP5},\ref{2.5}) we can write
\begin{equation}
\hat{g}(\vec{\tau}) = \hat{g}_s(\vec{\tau}) +
\hat{g}_{so}(\vec{\tau}) \vec{\sigma} [\vec{k} \times 2 \pi
\vec{\tau}] + g_p (\vec{\tau}) \vec{\sigma} \vec{\nu}_{1 \tau} +
g_t(\vec{\tau})\vec{\sigma} \vec{\nu}_{2 \tau}
\end{equation}
where $g_p (\vec{\tau})$ is proportional to the P-violating
scattering amplitude, $g_t(\vec{\tau})$ is proportional to the
T-violating scattering, $\vec{\nu}_{2 \tau}= \vec{\tau}/\tau$.
From (\ref{JP5}) we also have that the amplitudes $g_{\alpha}
(\vec{\tau})$ ($\alpha$ denotes $s$, $so$, $p$, $t$) can be
represented in the form
\begin{equation}
\hat{g}_{\alpha}^{({\tau})} = g_{1 \alpha}(\vec{\tau}) - i g_{2
\alpha}(\vec{\tau})
\end{equation}
where $g_{1 \alpha}(\vec{\tau})= g_{1 \alpha}(-\vec{\tau})$ and
$g_{2 \alpha}(\vec{\tau})= - g_{2 \alpha}(-\vec{\tau})$.

%---

\begin{eqnarray}
\label{JP10+}  \hat{g}_{\alpha}(\vec{\tau}) & = & -
\frac{2m}{\hbar^2 k_0^2} \hat{U}_{eff}^{(\alpha)}(\vec{\tau}) = -
\frac{2m}{\hbar^2 k_0^2}(-\frac{2 \pi \hbar^2}{m V_0})
\hat{F}_{\alpha}(\vec{\tau}) = \\
& = & \frac{4 \pi}{k_0^2} \frac{1}{V_0} \sum_j \hat{f}_{\alpha j}
(\vec{\tau}) e^{-W_j (\vec{\tau})} e^{-i 2 \pi \vec{\tau}
\vec{R}_j}, \nonumber \\
 \hat{g}_{1 \alpha} (\vec{\tau})  & = &  \frac{4 \pi}{k_0^2}
\frac{1}{V_0} \sum_j \hat{f}_{\alpha j} (\vec{\tau}) e^{-W_j
(\vec{\tau})} \cos 2 \pi \vec{\tau} \vec{R}_j, \nonumber \\
 \hat{g}_{2 \alpha} (\vec{\tau}) & = &  \frac{4 \pi}{k_0^2}
\frac{1}{V_0} \sum_j \hat{f}_{\alpha j} (\vec{\tau}) e^{-W_j
(\vec{\tau})} \sin 2 \pi \vec{\tau} \vec{R}_j, \nonumber \\
 \hat{g}_{1 t} (\vec{\tau})   & = &  g_{1T} (\vec{\tau})
\vec{\sigma} \vec{\nu}_{2 \vec{\tau}}, ~ \hat{g}_{2 t}
(\vec{\tau})  =  g_{2T}
(\vec{\tau}) \vec{\sigma} \vec{\nu}_{2 \vec{\tau}}, \nonumber \\
 \hat{f}_{Tj} (\vec{\tau})  & = &  f_{Tj} (\vec{\tau}) \vec{\sigma}
\vec{\nu}_{2 \vec{\tau}}= f_{Tj} (\vec{\tau}) \vec{\sigma}
\frac{\vec{\tau}}{|\vec{\tau}|}, \nonumber \\
g_{1T} (\vec{\tau})& = & \frac{4 \pi}{k_0^2 V_0} \sum_j f_{Tj}
(\vec{\tau}) e^{-W_j (\vec{\tau})} \cos 2 \pi \vec{\tau}
\vec{R}_j, \nonumber \\
g_{2T} (\vec{\tau}) & = & \frac{4 \pi}{k_0^2 V_0} \sum_j f_{Tj}
(\vec{\tau}) e^{-W_j (\vec{\tau})} \sin 2 \pi \vec{\tau} \vec{R}_j
\nonumber
\end{eqnarray}

When target nuclei are nonpolarized  $\hat g_s\left( \vec \tau
\right) = g_s\left( \vec \tau \right) $ does not depend on neutron
spin. Besides, for slow neutrons $g_{so}\ll g_s$. It allows to
write the contribution to caused by diffraction in the following
way

\begin{equation}
\delta \hat N=\frac{\hat g\left( -\vec \tau \right) \hat g\left(
\vec \tau \right) }{2\alpha _B}=\frac{g_s\left( -\vec \tau \right)
g_s\left( \vec \tau \right) }{2\alpha _B}+\frac 1{2\alpha
_B}g_s\left( -\vec \tau \right) \left[ \hat g_{so}\left( \vec \tau
\right) +\hat g_w\left( \vec \tau \right) \right] + \label{2.7}
\end{equation}

\[
+\frac 1{2\alpha _B}g_s\left( \vec \tau \right) \left[ \hat
g_{so}\left( -\vec \tau \right) +\hat g_w\left( -\vec \tau \right)
\right] .
\]

Let the target be nonpolarized. In this case T-noninvariant part
of $\hat g\left( \vec \tau \right) $ is \cite{VG,Bar}
\begin{equation}
\hat g_w^T\left( \vec \tau \right) =\frac{4\pi }{k^2V_0}\sum_j\hat
f_{jw}^T\left( \vec \tau \right) e^{-w_j\left( \vec \tau \right)
}e^{-i2\pi \vec \tau \vec R_j}, \label{2.8}
\end{equation}

\[
\hat f_{jw}^T\left( \vec \tau \right) =C_j^{^{\prime }}\left( \vec
\tau \right) \vec \sigma \frac{\vec \tau }\tau , C_j^{^{\prime
}}\left( \vec \tau \right) =ReC_j^{^{\prime }}+ImC_j^{^{\prime }}.
\]

From (\ref{2.7}),(\ref{2.8}) we see that the T-noninvariant
contribution to the refractive index in a nonpolarized crystals
may occur only in a non-center-symmetric crystals.

%---- addition from J.Phys.G

As a result, for the refractive index we can obtain
\begin{eqnarray}
\label{JP11} \hat{N} & = &1 + \frac{1}{2} g_s (0) + \frac{1}{2}
g_p (0)
\frac{\vec{\sigma} \vec{k}}{k}+ \\
& +& \sum_{r \ne 0} \frac{1}{2 \alpha_B(\vec \tau)} \left\{ -2i
[g_{1s} (\vec{\tau}) g_{2so} (\vec{\tau}) - g_{2s} (\vec{\tau})
g_{1so} (\vec{\tau})]  \vec{\sigma} [\vec{k} \times 2 \pi
\vec{\tau}] + \right. \nonumber \\
& + & 2 [g_{1s} (\vec{\tau}) g_{1p} (\vec{\tau}) + g_{2s}
(\vec{\tau})
g_{2p} (\vec{\tau})] \vec{\sigma} \vec{\nu}_{1s} - \nonumber \\
& - & 2i [g_{1s} (\vec{\tau}) g_{2T} (\vec{\tau}) - g_{2s}
(\vec{\tau}) g_{1T} (\vec{\tau})] \vec{\sigma} \vec{\nu}_{1 \tau}-
\nonumber \\
& - & 2i [g_{1so} (\vec{\tau}) g_{1p} (\vec{\tau}) + g_{2so}
(\vec{\tau}) g_{2p} (\vec{\tau})]  \vec{\sigma} [ [\vec{k} \times
2 \pi \vec{\tau}] \times \vec{\nu}_{1 \tau}] -
\nonumber \\
& - &  \left. 2i [g_{1so} (\vec{\tau}) g_{2T} (\vec{\tau}) -
g_{2so} (\vec{\tau}) g_{2T} (\vec{\tau})]  \vec{\sigma} [ [\vec{k}
\times 2 \pi \vec{\tau}] \times \vec{\nu}_{2 \tau}] \right\}
\nonumber
\end{eqnarray}
Spin-independent terms, which are proportional to $g^2(\tau)$ are
omitted because they do not influence the spin-dependent effects.

According to (\ref{JP11}) we can present $\hat{N}$ in the form
\begin{equation}
\hat{N}=1+\frac{1}{2}g_s (0) + \vec{\sigma} \vec{M} =
1+\frac{1}{2}g_s (0) + \vec{\sigma} \vec{M}^{\prime} + i
\vec{\sigma} \vec{M}^{\prime \prime} \label{JP12}
\end{equation}
where $\vec{M}^{\prime}$ and $\vec{M}^{\prime \prime}$ are the
real and imaginary parts of $\vec{M}$, respectively.
The term $\vec{\sigma} \vec{M}^{\prime}$ describes the neutron
spin rotation about the direction of $\vec{M}^{\prime}$, the term
$\vec{\sigma} \vec{M}^{\prime \prime}$ describes the spin
dichroism, i.e. the dependence of the neutron absorption on the
spin orientation (parallel or antiparallel to $\vec{M}^{\prime
\prime}$).
The spinor wavefunction of a neutron $\psi(l)$ can be represented
in the form
\begin{equation}
\psi(l) = e^{ik \hat{N}l} \psi(0) \label{JP13}
\end{equation}
where $\psi(0)$ is the spinor neutron wavefunction in vacuum and
$l$ is the neutron path length in a crystal.

The spin-dependent part of the phase in (\ref{JP13}) is only a
small part of the whole phase.
As a result, we can write (\ref{JP13}) as
\begin{equation}
\psi(l) \simeq [1+ i k \vec{\sigma} \vec{M} l ] e^{ik (1+
\frac{1}{2} g_s (0)l)} \psi(0) \label{JP14}
\end{equation}

From equation (\ref{JP14}) we obtain that the number of neutrons
passing through a crystal the path $l$ depends o the direction of
the neutron polarization vector $\vec{p}$:
\begin{equation}
J(\vec{p}) = J_0 e^{-\kappa l} [1-2 k (\vec{p} \vec{M}^{\prime
\prime} l)], \label{JP15}
\end{equation}
where $\kappa$ is the coefficient of the neutron absorption in a
crystal caused by the ordinary nuclear interaction, $|\vec{p}|=1$
for a completely polarized beam.
As we see, the intensity $J(\vec{p})$ depends on the orientation
of polarization vector $\vec{p}$ relative to $\vec{M}^{\prime
\prime}$.
The difference between the intensities $J_{\uparrow \uparrow}$
(for $\vec{p}$ parallel to $\vec{M}^{\prime \prime}$) and
$J_{\uparrow \downarrow}$ (for $\vec{p}$ antiparallel to
$\vec{M}^{\prime \prime}$) is
\begin{equation}
\frac{J_{\uparrow \uparrow} - J_{\uparrow \downarrow}}{J_{\uparrow
\uparrow} + J_{\uparrow \downarrow}} = -2 k p M^{\prime \prime} l.
 \label{JP16}
\end{equation}
It follows from (\ref{JP14}) that the polarization vector
$\vec{p}_1$ of the neutron passing through a crystal is
\begin{equation}
\vec{p}_1 = \frac{\langle \psi(l)| \vec{\sigma}| \psi(l)
\rangle}{\langle \psi(l)| \psi(l) \rangle} = \vec{p} + 2 k
[(\vec{p} \vec{M}^{\prime \prime}) \vec{p} - \vec{M}^{\prime
\prime}] l +2 k [\vec{p}\times \vec{M}^{\prime}] l.
 \label{JP17}
\end{equation}

The second term on the right hand side in (\ref{JP17}) describes
the appearance of neutron beam polarization due to the spin
dichroism effect.
This term exists even when the initial neutron polarization
$\vec{p}$ is zero.
In this case
\begin{equation}
\vec{p}_1 =  - 2 k  \vec{M}^{\prime \prime} l.
 \label{JP19}
\end{equation}
As we see, $\vec{p}_1$ is parallel to $\vec{M}^{\prime \prime}$.
From equation (\ref{JP11}) it follows that T-violating spin
rotation and spin dichroism phenomena appear only in crystals
without a center of symmetry.
We also see that the spin-orbital and P-violating, but
T-invariant, interactions contribute to spin rotation and
dichroism, too.
That is why the problem of separating T-violation effects from the
background of T-invariant phenomena arises.
%-----

Detailed analysis of possibility to distinguish different
contributions was done in \cite{VG,Bar}.
Let us consider more intently the term, for which
$\vec{k}^{\prime} = \vec{k} + 2 \pi \vec{\tau}_1 \simeq -\vec{k}$
i.e. addition to the refraction index from backward Bragg
reflection.
For this term the spin-orbital and P-violating, but T-invariant,
interactions are zero.

Suppose also that for the certain neutron energy Bragg parameter
$\alpha_B(\tau_1)$ for this term is $\alpha_B(\tau_1)\ll
\alpha_B(\tau)$, where $\tau_1 \ne \tau$.

According to (\ref{JP17}) the neutron polarization vector
$\vec{p}_1$ rotates in a crystal about the vector
$\vec{M}^{\prime}$.

As a result, the angle of particle spin rotation is
\begin{equation}
\vartheta = 2 k \vec{M}^{\prime} l = k Re(N_+  - N_-) L =
\vartheta_P + \vartheta_T.
 \label{JP18}
\end{equation}
where $\vartheta_P$ is the angle of spin rotation due to P-odd
T-even interactions of neutrons with nuclei described by the term
proportional to $\vec{\sigma} \frac{\vec{k}}{k}$ (see
(\ref{JP12})), $\vartheta_T$ is the angle of spin rotation caused
by P,T-odd interaction of neutrons with nuclei.
This rotation appears only due to diffraction.
The T-odd part of the scattering amplitude $f_T (\vec{\tau})$ sums
up two contributions: one caused by interaction of the nuclear
electric dipole moment with the electric field of nucleus
$f_{EDM}(\vec{\tau})$ and another appearing due to T-odd nuclear
interaction of the neutron with the nucleus $f_{TN} (\vec{\tau})$,
i.e. $f_T (\vec{\tau}) = f_{EDM}(\vec{\tau})+ f_{TN}
(\vec{\tau})$.
Therefore, the angle of rotation
\begin{equation}
\vartheta = \vartheta_{P} + \vartheta_{EDM} + \vartheta_{TN}.
 \label{+1}
\end{equation}
The term $\vartheta_P$ does not depend on $\alpha_B$ that makes
possible to distinguish it.
It also can be separated by the method considered in
\cite{LIYAF,LIYAF1}, which is based on double pass of the beam
through the crystal in forward and backward direction.

Let us consider the T-odd angle of rotation
\begin{equation}
\vartheta_{T} = \vartheta_{EDM} + \vartheta_{TN} = k Re (- \frac{2
i}{\alpha_B(\vec{\tau})} [g_{1s} (\vec{\tau}) g_{2T} (\vec{\tau})
- g_{2s}(\vec{\tau}) g_{1T} (\vec{\tau})]) \frac{\tau_z}{\tau}l.
 \label{+2}
\end{equation}
When the axis $z$ is directed along $\vec{k}$, then $\tau_z =
-\tau$.
Therefore,
\begin{eqnarray}
\label{+3} & & \vartheta_{T} = \vartheta_{EDM} + \vartheta_{TN} =
\frac{2 k}{\alpha_B} Re \left( i \left[ g_{1s}(\vec{\tau}) g_{2T}
(\tau) -
g_{2s} (\vec{\tau}) g_{1T} (\vec{\tau}) \right] \right) l = \nonumber \\
& & = \frac{2 k}{\alpha_B} Re \left( i \left[ g_{1s}(\vec{\tau})
g_{2 EDM} (\vec{\tau}) - g_{2s} (\vec{\tau}) g_{1 EDM}
(\vec{\tau}) \right] \right) l + \nonumber \\
& & +\frac{2 k}{\alpha_B} Re \left( i \left[ g_{1s}(\vec{\tau})
g_{2 TN} (\vec{\tau}) - g_{2s} (\vec{\tau}) g_{1 TN}
(\vec{\tau})\right] \right) l
 \end{eqnarray}

As it can be seen, measurement of the scattering angle
$\vartheta_T$ provides to find $g_t$ and, in its turn, to find sum
of contributions caused by EDM and P,T-odd interactions.
All this gives the limit for the scattering amplitude
$f_T(\vec{\tau}) = f_{EDM} +f_{TN}$, where the amplitude $f_{EDM}$
deals with the neutron EDM and $f_{TN}$ is determined by the
constant of the T-odd interaction.

The expression for the angle of rotation can be written as:
\begin{eqnarray}
 \label{+4}
&  & \vartheta_{T}  =  \frac{2 k}{\alpha_B(\tau)} \times \\
& & \times Re \left\{ i \left[ \frac{4 \pi}{k^2 V_0} \left( \sum_j
f_{sj}(\tau) e^{-W_j(\tau) \cos 2 \pi \vec{\tau} \vec{R}_j}
\right)
\frac{4 \pi}{k^2 V_0} \left( \sum_j f_{Tj}(\tau) e^{-W_j(\tau)
\sin 2 \pi \vec{\tau} \vec{R}_j} \right) - \right. \right. \nonumber \\
& &   - \left. \left. \frac{4 \pi}{k^2 V_0} \left( \sum_j
f_{sj}(\tau) e^{-W_j(\tau) \cos 2 \pi \vec{\tau} \vec{R}_j}
\right)
\frac{4 \pi}{k^2 V_0} \left( \sum_j f_{Tj}(\tau) e^{-W_j(\tau)
\sin 2 \pi \vec{\tau} \vec{R}_j} \right) \right] \right\}L
\nonumber
\end{eqnarray}

When the energy of interaction is far from resonances then the
potential scattering prevails.
In this case $Re f_s \gg Im f_s$ and the T-odd scattering
amplitude $f_T$ is purely imaginary.
Therefore, the expression in (\ref{+4}) is purely real.

In this case from (\ref{+4}) it follows that the angle
$\vartheta_t$ can be expressed as:
\begin{equation}
\vartheta_{T} \simeq k Re (N_s-1) L \frac{Re(N_s-1)}{\alpha_B}
\frac{Re \, f_T}{Im \, f_s}
 \label{+5}
\end{equation}

From (\ref{+5}) it follows that the effect of P-,T-odd spin
rotation is maximal when the value of $(N_s-1)$ is maximal.
The similar is also right for dichroism.

In particular, this is the reason for neutrons with the energy
close to the range of P-resonances to be interesting for study
\cite{VG,Bar}.
In this range both the amplitude $f_{TN}$  and the amplitude
$f_{EDM}$, which is caused by interaction of the electric dipole
moment of neutron with the nuclear electric field, demonstrate
resonant behavior.

Thus in the experiments for study of spin rotation and spin
dichroism  for neutrons in non-center-symmetric crystals both
contributions to the T,P-odd amplitude are considered:
contribution from the EDM and that from T,P-odd nucleon-nucleon
interaction.
It is interesting that for neutrons with the energy within the
thermal range and far from resonances the contribution from EDM
plays the major role in
 (\ref{+5}) \cite{Cherkas}.

This work is carried out within the joint grant of Belarusian
Republican Fund for Fundamental Research and Russian Fundamental
Research Fund $\#\Phi06$P-074.

\end{document}